\documentclass[aps,prl,reprint,showpacs,superscriptaddress,10pt]{revtex4-1}
\usepackage{amsmath,amssymb,amsfonts,graphicx,bm,longtable,dcolumn,color}
\usepackage{mathrsfs}
\usepackage[utf8]{inputenc}
\usepackage{textcomp}
\usepackage[T1]{fontenc}
\usepackage{lmodern}

\begin{document}

\title{Radiative heat shuttling}

\author{Ivan Latella}
\affiliation{Department of Mechanical Engineering, Universit\'{e} de Sherbrooke, Sherbrooke, PQ J1K 2R1, Canada}

\author{Riccardo Messina}
\affiliation{Laboratoire Charles Fabry, UMR 8501, Institut d'Optique, CNRS, Universit\'{e} Paris-Saclay, 2 Avenue Augustin Fresnel, 91127 Palaiseau Cedex, France}

\author{J. Miguel Rubi}
\affiliation{Secció de Física Estadística i Interdisciplinària - Departament de Física de la Matèria Condensada, Facultat de Física, Universitat de Barcelona, Martí i Franquès 1, 08028 Barcelona, Spain}

\author{Philippe Ben-Abdallah}
\email{pba@institutoptique.fr}
\affiliation{Department of Mechanical Engineering, Universit\'{e} de Sherbrooke, Sherbrooke, PQ J1K 2R1, Canada}
\affiliation{Laboratoire Charles Fabry, UMR 8501, Institut d'Optique, CNRS, Universit\'{e} Paris-Saclay, 2 Avenue Augustin Fresnel, 91127 Palaiseau Cedex, France}

\begin{abstract}
We demonstrate the existence of a shuttling effect for the radiative heat flux exchanged between two bodies separated by a vacuum gap when the chemical potential of photons or the temperature difference is modulated. We show that this modulation typically gives rise to a supplementary flux which superimposes to the flux produced by the mean gradient, enhancing the heat exchange. When the system displays a negative differential thermal resistance, however, the radiative shuttling contributes to insulate the two bodies from each other. These results pave the way for a novel strategy for an active management of radiative heat exchanges in nonequilibrium systems. 
\end{abstract}

\maketitle

Understanding the basic mechanisms which drive heat transfer in nonequilibrium systems is a fundamental goal in statistical physics. The interest in controlling heat exchanges at small scales is of primary importance for the development of a wide variety of technologies. In order to obtain such a control with a richness and flexibility comparable with the one available in electronics, phononic circuits have been proposed during the last decade to manipulate heat flows carried by phonons in the same manner as the flow of electrons is controlled in electric circuits~\cite{Casati1,BaowenLiEtAl2012,BaowenLi2}. 
Recently, to overcome the problems linked to the relatively small speed of acoustic phonons and to the presence of strong Kapitza resistances in these networks, photonic analogs have been proposed to control heat carried by radiation~\cite{Otey,Basu,Fan,PBA_APL,PBA_PRL2014,Kubytskyi,PBAEtAl2015,PBA2016,PBAlogic,ItoEtAl2015,ItoEtAl2016,Moncada,pba_Hall,Fan2,Latella_PRL2017,Cuevas}.

In this Letter, we explore the temporal evolution of radiative heat transfer between two bodies across a separation gap by modulating one of the intensive quantities which are responsible for heat exchanges. By studying the heat transfer between two dielectrics with an oscillating temperature difference and between a semiconductor and a dielectric with an oscillating photon chemical potential difference, we prove the existence of a radiative heat shuttling, a supplementary flux superimposed to the one produced by the mean gradient. We demonstrate that this effect is intrinsically related to the nonlinearity of radiative exchanges. When the optical properties of the media do not change during the time evolution of either the temperature or the chemical potential, the shuttling amplifies the transfer between the two bodies. On the contrary, if those properties change, we show that the shuttling can insulate the two bodies from each other when the system supports a negative differential thermal resistance. 

\begin{figure}
\includegraphics[scale=1]{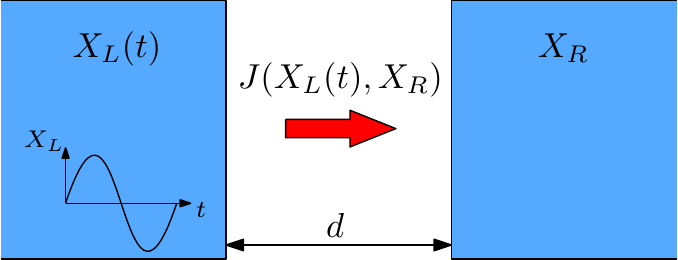}
\caption{Sketch of a device displaying radiative heat shuttling. The system consist of two bodies separated by a vacuum gap of thickness $d$ that exchange an energy flux $ J $ by means of thermal radiation. The intensive property $ X_L ( t ) $ of the left body is modulated in time. }
\label{fig1}
\end{figure}

To illustrate the radiative heat shuttling, we consider two different bodies, labeled $L$ and $R$, separated by a vacuum gap of thickness $d$, and emitting thermal radiation at temperatures $ T_i $ and chemical potentials $ \mu_i $ for $ i = L, R $. As it is well known, if the body is a semiconductor subject to an applied voltage $V_i$, the chemical potential of the radiated photons is $\mu_i=qV_i$ at frequencies above the gap frequency of the material and zero otherwise, $q$ being the elementary charge~\cite{Wurfel,Henry,Chen}. We then consider that the following oscillating gradient as a function of the time $ t $ is applied between the two bodies:
\begin{equation}
\begin{split}
X_L ( t ) & = x_L + \delta X \sin( \Omega t ) , \\
X_R & = x_R , 
\end{split} 
\label{Eq:X}
\end{equation}
where 
$ X_i = T_i $ with fixed $ \mu_i $ or $ X_i = \mu_i $ with fixed $ T_i $, and $ \Omega = 2 \pi / \tau $ is the modulation frequency. Here $ x_L $ and $ x_R $ (with $ x_L \geq x_R $) are some reference values of temperatures or chemical potentials and $\delta X$ is the amplitude of the oscillation in the left reservoir. In order to avoid dealing with the coupling between conduction and radiation, we assume here that the oscillation period $\tau$ is large enough compared to the thermalization time. As we will see, even for a vanishing time average of $ \Delta X ( t ) = X_L ( t ) - X_R $, defined as $ \Delta \bar{ X } = 1 / \tau \int _0 ^\tau \Delta X ( t ) \, dt = x_L - x_R $, the net energy flux transferred between the slabs does not vanish.

The heat flux exchanged between the two bodies reads~\cite{Polder,Pendry}
\begin{equation}
J ( X_L(t), X_R ) = \int_0^\infty \frac{ d \omega }{ 2 \pi } \Theta_{ LR } K_{ LR } ,
\label{Eq:heat_flux}
\end{equation}
where $\omega$ is the frequency of the electromagnetic field, $ K_{ LR } ( \omega, X_L(t), X_R ) > 0 $ represents the effective number of modes participating in the transfer (given by the sum of the transmission coefficients over the two polarizations integrated over all possible parallel wavectors), and $ \Theta_{ LR } ( \omega, X_L(t), X_R ) = \Theta_L ( \omega, X_L(t) )- \Theta_R ( \omega, X_R )$ denotes the difference of energy of Planck oscillators defined as
\begin{equation}
\Theta_i ( \omega, X_i ) =
\begin{cases} 
\hbar \omega \left[ e^{ (\hbar \omega) / ( k_B T_i ) }-1 \right]^{ -1 }, & X_i = T_i, \\
\hbar \omega \left[ e^{ (\hbar \omega - \mu_i ) / ( k_B T_i ) }-1 \right]^{ -1 }, & X_i = \mu_i,
\end{cases}
\label{Eq:Planck}
\end{equation}
$ \hbar $ and $ k_B$ being the reduced Planck constant and the Boltzmann constant, respectively.
Expanding the spectral flux $ \varphi_{ LR } = \Theta_{ LR } K_{ LR } $ around $ X_L = x_L $ yields
\begin{equation}
\begin{split}
\varphi_{ LR } & = \varphi_{ LR } ^0 + \delta X \sin ( \Omega t) \left. \frac{ \partial \varphi_{ LR } }{ \partial X_L} \right|_{X_L = x_L}\\
& + \frac{ \delta X^2 }{ 2 } \sin^2( \Omega t) \left. \frac{ \partial^2 \varphi_{ LR } }{ \partial X_L^2 } \right|_{ X_L = x_L } + O( \delta X^3 ),
\label{Eq:tailor}
\end{split}
\end{equation}
where $ \varphi_{ LR } ^0 $ is the spectral flux without oscillation of the corresponding intensive quantity.
It turns out that the average heat flux reads
\begin{equation}
\bar{J} = \frac{1}{\tau} \int_{0}^\tau \! J(t) dt \simeq J_{ LR } ^0 + \frac{ \delta X^2 }{ 4 } \left. \frac{ \partial^2 J }{ \partial X_L^2 }\right|_{X_L=x_L},
\label{Eq:m_flux}
\end{equation}
$ J_{ LR } ^0 $ being here the heat flux evaluated at $ X_L = x_L $ and $ X_R = x_R $.

The shuttling of heat is related to the local curvature of the flux with respect to the intensive quantity $X_L$ and therefore to the sign of
\begin{equation}
 J'' =\frac{ \partial^2 J }{ \partial X_L^2 }=\int_{0}^\infty\! \frac{ d\omega }{ 2\pi } \left( \Theta''_L K_{ LR } + 2 \Theta'_L K'_{ LR } + \Theta_{ LR } K''_{ LR } \right)
\label{Eq:dervative}
\end{equation}
at $ X_L = x_L $, where primes denote derivative with respect to $ X_L $.
When $ J $ is convex (i.e. $ J'' > 0 $), the shuttling produces an increase of the total flux. Such a curvature always exists when the optical properties of the involved materials do not depend on either the temperature or chemical potential. It is noteworthy that under these conditions, the shuttling always amplifies the heat transfer. Indeed, in this case the average flux reduces to
\begin{equation}
\bar{ J } \simeq J_{ LR } ^0 + \frac{ \delta X^2 }{ 4 } \int_{0}^\infty\! \frac{ d\omega }{ 2\pi } \Theta''_L K_{ LR }, 
\end{equation}
and it is easy to verify that $ \Theta''_L > 0 $ over the whole spectral range so that the shuttling flux is positive and therefore $ \bar{ J } \geq J^0_{ LR }$. On the contrary, if $ J $ is locally concave (i.e. $ J'' < 0 $), the shuttling effect tends to inhibit the flux. We will see below that this situation can occur when and only when the system possesses a negative differential resistance. As shown in previous works~\cite{PBA_PRL2014}, the presence of such an anomalous resistance requires a dependence of optical properties with respect to $X_L$. With this in mind, it can be seen from Eq.~(\ref{Eq:dervative}) that a negative shuttling is plausible in a simple situation. Choosing $ x_L = x_R = x_0 $ as a background intensive quantity, the zeroth-order contribution to the flux $ J_{ LR } ^0 $ and the difference of energies $ \Theta_{ LR } $ cancel out when evaluated at $ X_L = x_0 $. Thus, the time average flux takes the form
\begin{equation}
\bar{ J } \simeq \frac{ \delta X^2 }{ 4 } \int_{0}^\infty\! \frac{ d\omega }{ 2\pi } \left( \Theta''_L K_{ LR } + 2 \Theta'_L K'_{ LR } \right). 
\label{Eq:negative_flux}
\end{equation}
It is straightforward to notice that since $ \Theta'_L > 0 $, this flux can be negative provided the number of modes $ K_{ LR} $ decreases fastly enough with $ X_L $.

\begin{figure}
\includegraphics[scale=1]{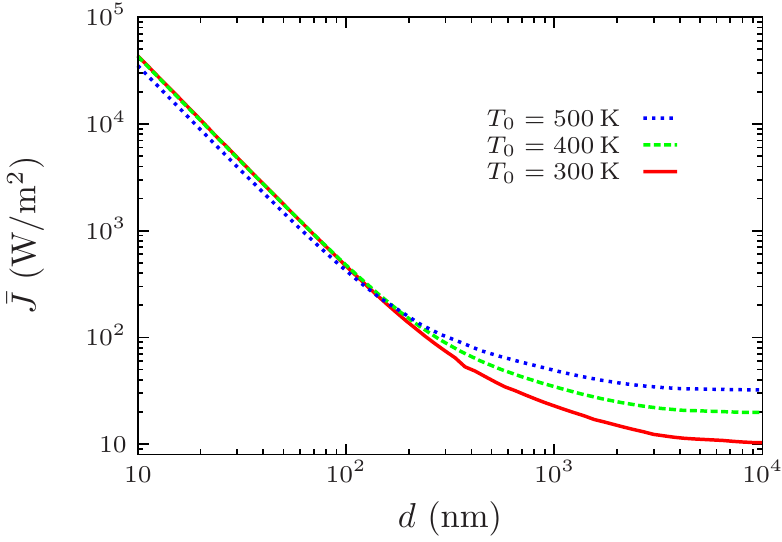}
\caption{Radiative heat shuttling between two thick SiO$_2$ slabs when $ T_L ( t ) =T_0+\delta T \sin(\Omega t)$ and $ T_R ( t ) =T_0$. The time average flux $ \bar{ J } $ is shown as a function of the separation distance $ d $ for different values of the background temperature $ T_0 $. The amplitude of the oscillations is $\delta T = 30\,$K and the modulation frequency $ \Omega$  is small with respect to the inverse of the thermalization time.}
\label{fig2}
\end{figure}

To exemplify the radiative heat shuttling effect in a particular situation, we first consider the heat transfer between two silica (SiO$_2$) samples whose permittivity modeling the thermal emission is taken from~\cite{Palik}. We assume a periodically time-varying temperature $ T_L ( t ) = T_0 + \delta T \sin( \Omega t ) $ on the left sample and a fixed temperature $ T_R ( t ) = T_0 $ on the right sample (see Fig.~\ref{fig1}). Hence both the average temperature difference $ \Delta \bar{ T } $ and $ J_{ LR } ^0 $ vanish. The average flux $ \bar{ J } $ is plotted in Fig.~\ref{fig2} as a function of the separation distance $d$ for several values of the background temperature $ T_0 $ with a fixed modulation amplitude $ \delta T = 30\,$K. We remark that in the numerical examples hereafter, $ \bar{ J } $ is computed from the time average of the exact expression (\ref{Eq:heat_flux}). As typically happens with the energy flux for polar materials as SiO$_2$, the average $ \bar{ J } $ scales as $ d^{ -2 } $ in the near field, i.e. at small separations, where the energy transfer is mainly driven by the contribution of surface phonon polaritons supported by the material. Moreover, we observe that $ \bar{ J } $ weakly increases in this regime as the temperature $ T_0 $ decreases, which can be understood as a consequence of the overlap in the spectrum between $ \Theta''_L $ and the peak in $ K_{ LR } $ accounting for the resonant mode. On the other hand, in the far field $ \bar{ J } $ approaches a constant value which as expected increases with $ T_0 $. 

\begin{figure}
\includegraphics[scale=1]{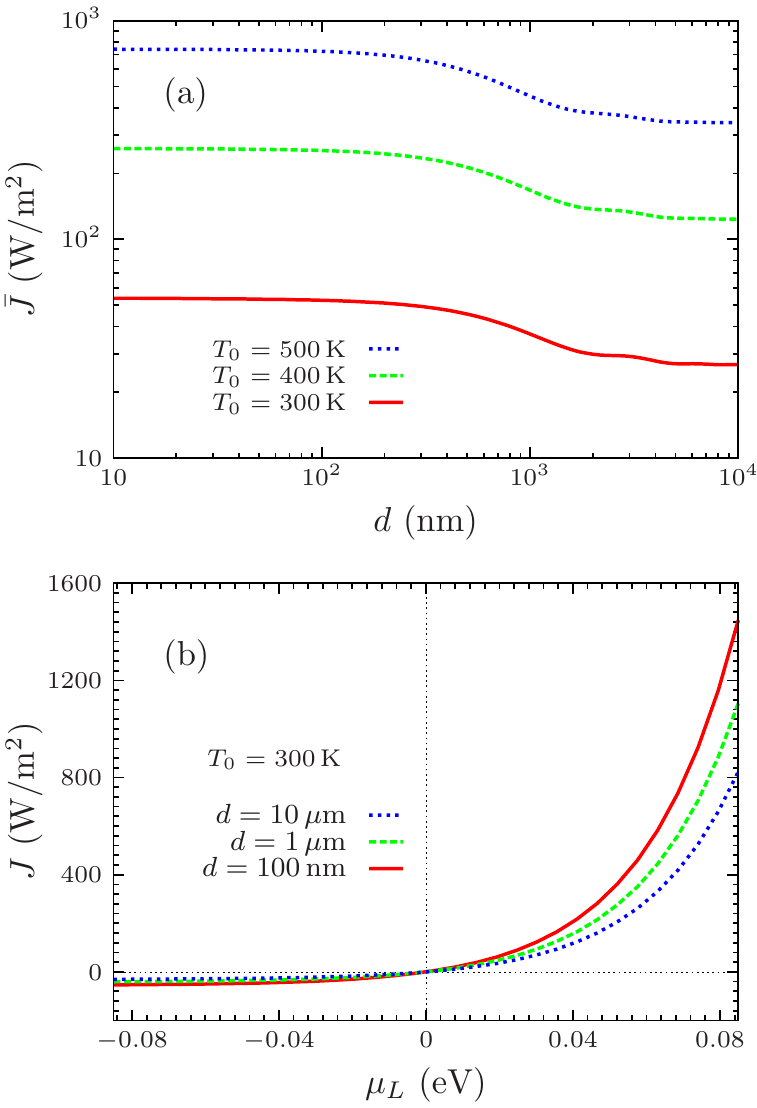}
\caption{Radiative heat shuttling between a direct-gap semiconductor (InSb) and a dielectric (SiO$_2$). The two materials are thermalized at the same temperature $ T_0 $ and the semiconductor is subject to an applied oscillating voltage $ V_L ( t ) $. (a) Time-averaged heat flux $ \bar{ J } $ exchanged between the bodies as a function of the separation distance $ d $. (b) Corresponding flux $ J $ as a function of the chemical potential $ \mu_L $ of the photons emitted by the semiconductor.}
\label{fig3}
\end{figure}

As a second example, we consider the transfer between a narrow-band gap semiconductor (left body) whose electrons and holes are excited by an external bias voltage $ V_L ( t ) $ and a passive dielectric slab (right body). The temperatures of the bodies are maintained fixed by coupling them with an external thermostat at temperature $T_0$, so that $ T_L = T_R = T_0$. For clarity reasons we assume a periodic modulation $ V_L ( t ) = \delta V \sin( \Omega t ) $ of the applied bias voltage. From here, the chemical potential of photons emitted by the semiconductor is $ \mu_L ( t ) = q\, \delta V \sin( \Omega t ) $ above the gap frequency $ \omega_g = E_g / \hbar $ and $ \mu_L ( t ) = 0 $ below this frequency, where $ E_g $ is the gap energy. In our example, the left body is made of indium antimonide (InSb) whose permittivity is given in~\cite{Forouhi} by fitting experimental data and for the gap energy we take $ E_g = 0.17\,$eV.  As before, we assume the right body to be made of SiO$_2$ and therefore, the chemical potential of the radiation emitted by this material is $ \mu_R = 0 $. Since the two slabs are thermalized at the same temperature, we remark that the energy difference $ \Theta_{ LR } $ vanishes below the gap frequency and thus, the spectrum of the energy flux is different from zero only in the frequency range above $ \omega_g $. The resulting flux $ \bar{ J } $ for this case is shown in Fig.~\ref{fig3}(a) as a function of $ d $ for several temperatures $ T_0 $, where the amplitude of the oscillation is taken as $ \delta \mu = q \delta V = E_g / 2$. We  observe that  $ \bar{ J } $ increases when the background temperature $ T_0 $ increases. Furthermore, we highlight that in the temporal evolution, $ J $ becomes negative when the chemical potential is negative. However, in this situation the magnitude in modulus of $ J $ is smaller than the values reached when $ J > 0 $, making the time average a positive quantity. This flux is plotted in Fig.~\ref{fig3}(b) as a function of the chemical potential $ \mu_L $. We clearly see  a diode-like behavior with a strong asymmetry (rectification coefficient larger than $95\%$). It immediately follows from this asymmetric behavior that by using a temporal modulation of chemical potential with a nonvanishing average value, it is possible to change the sign of the shuttling flux. 

\begin{figure}
\includegraphics[scale=1]{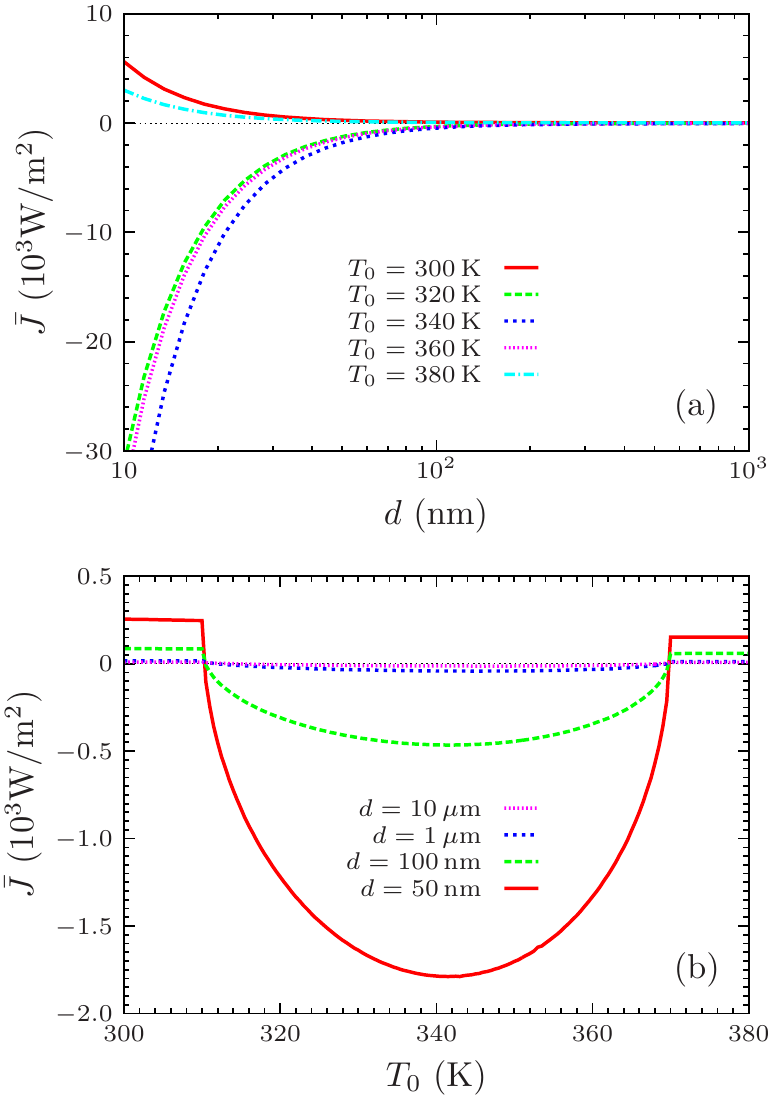}
\caption{(a) Time-averaged flux $ \bar{ J } $ as a function of the separation $d$ between a VO$_2$ slab and a sample of SiO$_2$. The temperature $ T_L ( t ) $ of the VO$_2$ slab is periodically modulated around $ T_0 $ with an amplitude $ \delta T = 30\,$K, whereas the temperature of the other body is fixed at $ T_R = T_0 $. (b) Average flux  $ \bar{ J } $ with respect to $ T_0 $.}
\label{fig4}
\end{figure}

In the previous examples, a net heat flux is always transferred from an active body to a passive body with a sinusoidal modulation of the intensive property. We now consider the heat exchange between a material characterized by an insulator-metal transition (IMT), which is able to undergo an important change in its optical properties through a small change of its temperature around a critical temperature $ T_c $, and a dielectric material whose optical properties do not depend on the temperature. With this example, we show that a heat flux can be extracted from the passive body, i.e. $ \bar{ J } < 0 $, as already pointed out in our previous discussion. For the IMT material, we take vanadium dioxide (VO$_2$) which undergoes a first-order transition (Mott transition~\cite{Mott}) from a low-temperature insulating phase to a high-temperature metallic phase close to room temperature ($ T_c = 340\,$K). The optical properties of VO$_2$ in the two phases are given in~\cite{Barker}. The temperature $ T_L ( t ) $ of the VO$_2$ slab is periodically modulated around a temperature $ T_0 $ close to $ T_c $, so that the medium undergoes a transition during this modulation. The oscillations amplitude  is set, as previously, at $ \delta T = 30\,$K.  For the second body, we use once again a SiO$_2$ sample. During the modulation process, the temperature of the SiO$_2$ sample is held fixed at $ T_R ( t ) = T_0 $ while the two photon chemical potentials are supposed to vanish  (i.e. $ \mu_L = \mu_R = 0$). The average flux $ \bar{ J } $  plotted in Fig.~\ref{fig4}(a) with respect to the separation distance $ d $ shows two radically opposite behaviors. On the one hand, we observe that $ \bar{ J } $  is positive when $ T_0 $ is sufficiently far from the critical temperature $ T_c$ so that the VO$_2$ slab does not undergo a phase transition during its temperature modulation. On the other hand, $ \bar{ J } $ is negative when $ T_c $ is reached during the modulation. This behavior is corroborated in Fig.~\ref{fig4}(b) where $ \bar{ J } $ is  plotted with respect to $ T_0 $ for different separations $ d $. Here we clearly observe that $ \bar{ J } < 0 $ when $ T_0 $ is chosen in the range  between $ T_c - \delta T $ and $ T_c + \delta T $. According to Eq.~\eqref{Eq:negative_flux}, this negative average flux, acting as a heat-pumping mechanism, is a direct consequence of the drastic reduction of the number of modes contributing to the energy transfer during the temperature modulation of the IMT material. This reduction is related to the mismatch between the emission spectra of the IMT material and the dielectric beyond the critical temperature, which gives rise to a radiative negative differential thermal resistance between the two bodies. In the presence of an oscillating temperature difference, this resistance --and therefore the shuttling effect-- tends to insulate the two bodies from each other. Furthermore, as shown in  Fig.~\ref{fig4}(b), it is worth noting that to maximize this effect $ T_0 $ must be equal to $ T_c $.

So far, we have only considered sinusoidal modulations of the intensive quantity $ X_L (t) $ under scrutiny. We also would like to remark that similar effects can be observed with periodic modulations [i.e. $ X_L ( t ) = X_L ( t + \tau ) $] which are neither odd nor even functions or with random modulations (in this case $\tau\rightarrow\infty$). Indeed, in such situations we have $ \bar{ X }_L \neq x_L$, so that 
\begin{equation}
\bar{ J } = J^0_{ LR } + ( \bar{ X }_L - x_L ) \left. \frac{ \partial J }{ \partial X_L }\right|_{ X_L = x_L } + O \bigl( ( \bar{ X }_L - x_L )^2 \bigr), 
\label{Eq:m_flux2}
\end{equation}
where
\begin{equation}
\frac{ \partial J }{ \partial X_L } = \int_{0}^\infty\! \frac{ d \omega }{ 2 \pi } \left( \Theta'_L K_{ LR } + \Theta_{ LR } K'_{ LR } \right).
\label{Eq:dervative2}
\end{equation}
Since $ \Theta'_L $ is always a positive function, it is in principle possible to obtain a pumping effect (i.e. $  \bar{ J } - J^0_{ LR } <0 $) when $ \bar{ X }_L - x_L > 0 $ (always assuming $ x_L > x_R $), provided the gradient $ \partial J / \partial X_L $ in Eq.~\eqref{Eq:dervative2} is negative. However, we see that such a pumping also requires the presence of a local negative differential thermal resistance.

Beside its fundamental interest, the radiative heat shuttling effect could play an important role in the development of Brownian heat engines driven by photon tunneling at the nanoscale, in which the temperature fluctuations $\delta T$ vary as $ T_0/N^{1/2}$  around its average temperature $T_0$ and they can be relatively large when the number $N$ of atoms is  small. Hence, at ambient temperature $\delta T\sim10 \:K$ for a particle of one hundred atoms. Furthermore, using temporal modulations of local temperatures or chemical potentials, the shuttling of energy in systems composed by nanoemitters coupled in the near field with mechanical oscillators could give a supplementary flexibility to control heat exchanges, a relevant feature for nano-electromechanical systems technology. Another interest of radiative heat flux control with temperature or chemical potential modulation concerns energy management applications, since this mechanism could be more efficient than that with a static control from an external thermostat or current generator. Indeed, in the scenario of heat-flux control using, for instance, a sinusoidal modulation of the photon chemical potential, the power dissipated in the system is $\mathscr{P}=\frac{1}{2}U_\text{max}\,I_\text{max}\,\cos(\varphi)$, where $\varphi$ denotes the dephasing between current and voltage. Therefore, by controlling the phase shift it is in principle possible to make the dissipated power arbitrarily small, contrarily to the case with absence of modulation. 

In summary, we have demonstrated the existence of a shuttling of radiative heat between two bodies under nonequilibrium conditions. We have seen that this effect can be induced either by modulating the temperature or the chemical potential of the radiation in the case of semiconductors. Moreover, this phenomenon can be used to amplify or to reduce the radiative heat transfer between the interacting bodies. In order to observe the insulating behavior, we have demonstrated that  a negative differential thermal resistance must exist between the two bodies. Finally, we have also shown that a temporal modulation of the chemical potential of the radiation emitted by a semiconductor could be used to efficiently amplify the energy transfer between the materials over a wide temperature range.

\begin{acknowledgments}
The authors are grateful to P. Santhanam and K. Chen for very fruitful exchanges on the optical properties of semiconductors under a bias voltage. 
\end{acknowledgments}

\end{document}